\documentstyle[12pt,epsf]{article}
\newcommand{\rf}[1]{(\ref{#1})}
\newcommand{\beq}{\begin{equation}}
\newcommand{\eeq}{\end{equation}}
\newcommand{\bdm}{\begin{displaymath}}
\newcommand{\edm}{\end{displaymath}}
\newcommand{\bea}{\begin{eqnarray}}
\newcommand{\eea}{\end{eqnarray}}
\newcommand{\nn}{\nonumber \\}
\newcommand{\al}{\alpha}
\newcommand{\g}{\gamma}
\newcommand{\del}{\delta}
\newcommand{\Del}{\Delta}

\newcommand{\ve}{\varepsilon}
\newcommand{\mn}{\mu\nu}

\newcommand{\pa}{\partial}

\newcommand{\lam}{\lambda}

\newcommand{\bfx}{\bf{x}}

\newcommand{\oB}{\overline{B}}
\newcommand{\oI}{\overline{I}}
\newcommand{\oR}{\overline{R}}

\newcommand{\cO}{{\cal{O}}}

\newcommand{\G}{{\rm G}}

\newcommand{\seed}{B_{\rm seed}}
\newcommand{\gut}{\alpha_{\rm GUT}}

\def\gsim{\raise.3ex\hbox{$>$\kern-.75em\lower1ex\hbox{$\sim$}}}
\def\lsim{\raise.3ex\hbox{$<$\kern-.75em\lower1ex\hbox{$\sim$}}}

\begin{document}

 {\title
{\null\vskip-3truecm
{ \hskip10truecm {\small NORDITA-94/6 P\hfill }\vskip 0.1cm}
{\bf On primordial  magnetic fields$^*$}}
\author{
{\sc Kari Enqvist$^\dagger$} \\
{\sl Nordita} \\
{\sl Blegdamsvej 17, DK-2100 Copenhagen, Denmark}}
 \maketitle}

\vspace{1.5cm}
\begin{abstract}
\noindent
 A primordial magnetic field could be responsible for the observed magnetic
fields of the galaxies. One possibility is that such a primordial field
is generated at the electroweak phase transition because of the fluctuations
in the Higgs field gradients. I describe a statistical averaging procedure
which gives rise to a field of a correct magnitude. Another possibility,
where the Yang-Mills vacuum itself is ferrromagnetic, is also discussed.
\vspace{2cm}
\end{abstract}
\vfil
\footnoterule
{\small  $^*$Talk given at the NATO workshop {\sl Electroweak Physics
and the Early Universe}, Sintra, Portugal, March 1994.\\
$^\dagger$enqvist@nbivax.nbi.dk }
\thispagestyle{empty}
\newpage
\setcounter{page}{1}


\section{Introduction}
The very early universe is in notoriously short supply of observables
that would have survided until today. It is likely that the baryon
number of the universe and the   density perturbations are one.
Another, but a much more speculative possibility,
might be the magnetic fields of spiral
galaxies, the origin of which still largely remains a puzzle.
The nearby galaxies   have magnetic fields of the order
of $B\simeq 10^{-6}$ G \cite{dynamo}, which  can be
deduced from observations of the syncrotron radiation
put out by electrons travelling through the fields, assuming equipartition
of magnetic and particle energies\footnote{Equipartition may not to be
valid for certain irregular galaxies \cite{Wolfendale}.}.
Recently, a field of a similar magnitude has been observed also in a
object with z=0.395 \cite{bigz}.

The   model for galactic magnetic
fields most studied is the galactic dynamo \cite{brand},
where differential rotation and  turbulence
of the ionized gas
amplifies a  weak seed field by several orders of magnitude.
Not much is known about the seed field.
As the dynamo  growth time of the magnetic field cannot be smaller than the
galactic rotation period $\tau\simeq 3\times 10^8$ yrs, this gives
a lower limit of $\seed\gsim 10^{-19}$ G on a comoving
scale of the protogalaxy, about
100 kpc. In the Milky Way and the Andromeda Nebula the
dynamos appear to be rather weak and the growth time as long as
$\tau \simeq 10^9$ yrs which would imply that $\seed\gsim B\exp (-t/\tau)
\simeq 10^{-10}$ G. Moreover,
in the Milky Way  the magnetic field changes its
direction by about 180$^o$ between the Sagittarius and the Orion spiral
arms \cite{orion}, and it has been argued \cite{anvar}
 that such a reversal implies a stringent lower bound of
$\seed\gsim 10^{-7}$ G
on the seed field. As such a reversal has only been observed in the
Milky Way, it might not be a generic feature.

One interesting possibility is that the seed field
is truly primordial, with an origin that predates nucleosynthesis.
In that case the protogalaxy collapsed with a frozen-in magnetic field,
which enhanced the primordial cosmological field by a factor of $10^4$
\cite{Shukurov}.
Thus at the scale of 100 kpc
the dynamo mechanism requires a primordial field  somewhere in the ballpark of
of $10^{-18}$ G, with an uncertainty of a few orders of magnitude.

{}From a theoretical point of view, however, the generation of a sufficiently
large persistent
magnetic field in the early universe is rather difficult. There are
various attempts, relying on phase transitions such as the cosmic inflation
or the QCD phase transition
\cite{attempts}, but the field often  comes out to be too small to be of
cosmological interest. It has been suggested that a large
field might actually be generated at the electroweak phase transition
because of  random fluctuations in the
 Higgs field \cite{vachaspati}.
If one assumes a stochastic, uncorrelated distribution of the
Higgs field gradients,
or of the magnetic field itself, one finds \cite{eo2} today
at 100 kpc a root--mean--square field of the order of $10^{-19}$ G,
which could well serve as the origin of the seed field. This
positive result is based
on calculating the statistical averages along an arbitrary curve.
This is not the only possibility, but averaging over areas or volumes
would produce a field far too small to be of relevance for the dynamo
effect.

In Yang-Mills theories  there is also the possibility \cite{savvidy}
that the vacuum is
an analog of the ferromagnet with a non-zero background magnetic field.
This is a non-perturbative effect, and the resulting field is typically
very small. If one is willing, however, to go up all the way to the GUT
scale one finds that a typical GUT phase transition could have given
rise to a background field large enough to serve as the seed field \cite{eo3}.

\section{EW magnetic fields}
Electromagnetism first occurs when the standard electroweak
$SU(2) \otimes U(1)_Y$ theory is broken down to $U(1)_{\rm em}$.
It is therefore particularly attractive that Vachaspati \cite{vachaspati}
has explained the origin of a primordial field in terms of the
cosmological boundary condition that all physical quantities
should be uncorrelated over distances greater than the horizon
distance.
Since we start with the group $SU(2) \otimes U(1)_Y$ before the
electroweak phase transition, the resulting electromagnetic field
can be constructed in a way which is different from the usual
$F_{\mn} = \pa_\mu A_\nu - \pa_\nu A_\mu$.
The result is \cite{vachaspati}
\bea
F_{ij} & = & - i(V^\dagger_i V_j - V_i V^\dagger_j)~~, \nn
V_i & = & \frac{2}{|\phi|} \sqrt{\frac{\sin \theta}{g}}~
\pa_i \phi ~~,                                           \label{2}
\eea
where $\phi$ is the Higgs field.
At the electroweak phase transition the correlation length in the
broken phase is
$\sim 1/m_W$ (assuming that the Higgs mass is comparable to
$m_W$).
The field $F_{ij}$ is thus constant over a distance $\sim 1/m_W$, but
it varies in a random way over larger distances. Its variation is due to the
fact that the field $\phi$ makes a
random walk on the vacuum manifold of $\phi$.
The problem then is to estimate the field $F_{ij}$ over a length scale
$\sim N/m_W$.
If $N = 1$, it  then follows that on dimensional grounds $F_{ij} \sim
m_W^2 \sim 10^{24} $ G, with probably an uncertainty of $\pm 1$ in
the exponent.
For $N$ large, one should use an approriate statistical argument.
The issue is, which is the appropriate way to average over the random
fields.

Let us consider random fields walking around in space in a certain
number of steps.
Thus we replace the continuum by a lattice, where the points are
denoted by greek letters $\al, \ldots$.
I wish to estimate the magnetic field over a {\em linear} scale
(which at most is equal to the horizon scale).
Thus, let us consider a curve consisting of $N$ steps in the
lattice and  define the mean value
\beq
\oB = \frac{1}{N} \sum^N_{i=1} B^{\al_{i}}~~,          \label{6}
\eeq
where $B$ is a component of the magnetic field, and where the
lattice points $\al_i$ are on the curve.

Now this curve is arbitrary, and we could take any other curve.
Let us therefore define the average
$\langle  \ldots \rangle$, which averages over
curves spanning an $N^3$ lattice, i.e. over all space. Then, for example,
\beq
\langle  \oB \rangle = \frac{1}{N} \langle \sum^N_{i=1} B^{\al_{i}}
\rangle~~,                               \label{7}
\eeq
which means that for each curve with $N$ steps the mean value
$\oB$ is computed, and this is done for a set of curves which span an
$N^3$-lattice,
and the average is then computed.
Therefore $\langle \oB\rangle$ depends in general on $N$, but for simplicity of
notation we shall leave out the explicit reference to this dependence.
It should be  emphasized that the ensemble average \rf{7} takes into account
the field value at each lattice point, so that {\it the average is really over
the whole lattice volume} \cite{eo2}.

Similarly, one can define higher moments such as
\beq
\langle  \oB^2 \rangle = \frac{1}{N^2}
\sum^N_{i,j=1} \langle B^{\al_{i}} B^{\al_{j}}\rangle~~,  \label{8}
\eeq
together with quantities like
$
\langle ( \oB - \langle \oB\rangle )^2  \rangle~~.
$
Note that in  \rf{8}  the sum is over
curves of length $N$ steps of the non-local quantity
$\langle B^{\al_{i}} B^{\al_{j}} \rangle$.

In \cite{vachaspati} the stochastic variable was taken to be the Higgs field
itself which varies over the vacuum manifold.
Vachaspati argued that the gradients are of order
$1/ \sqrt{N}$, since $\phi$ makes a random walk on the vacuum manifold
with $\Del \phi \sim \sqrt{N}$, and since $\Del x \sim N$.
Thus $V_i$ is, in a root mean square sense, of the order
$1/ \sqrt{N}$, and hence $F_{ij}$ is of order $1/N$.
Taking further into account that the flux in a co-moving circular
contour is constant, the field must decrease like $1/ a(t)^2$, where
$a(t)$ is the scale factor.
Using the fact that in the early universe $a$ goes like the
inverse temperature, the field
was then estimated to behave like
$\langle F_{ij}\rangle_T~ \sim  {T^2}/{N}$.
For a scale of 100 kpc this leads to $\langle F_{ij}\rangle_{now} ~\sim
10^{-30} $ G, which is far too small
to explain the galactic fields (unless there exists some large scale
amplification mechanism). One should also point out that this assumption
presumes that the total magnetic flux through a given surface is a
stochastic sum of the fluxes through the unit cells of that surface,
or in other words, that the fluxes through two adjoining unit cells are
uncorrelated. Whether this is true or not is an open question.

It is however natural to assume that also the gradient vectors $V_i$ are
stochastic, as was done in \cite{eo2}.
This is because they directly specify whether there is a magnetic
field or not, whereas this is only true indirectly for the Higgs
field itself.
Also, the vectors $V_i$ are relevant for questions of alignment
between neighbouring domains.

\section{Random Higgs gradients}

Consider  the expression \rf{2} for the magnetic field in
terms of the Higgs gradients $V_i$.
It is convenient to split these fields into real and imaginary parts,
\beq
V_i ({\bfx}) = R_i ({\bfx}) + i I_i ({\bfx})~~,                \label{9}
\eeq
where $R_i$ and $I_i$ are real vectors.
Let us consider the system at a fixed time.
The cosmological boundary condition is then that $R_i$ and $I_i$ are
random fields.
Let us make the following assumptions:
\begin{itemize}
\item[(i)] The random fields have a Gaussian distribution.
Thus, the mean value of some quantity $Q$ is given by
\beq
\langle Q\rangle = \prod_{\al,i} \int \frac{d^3 R^\al_i}{D} \frac{d^3
I^\al_i}{D}
Q~e^{- \lam (R^\al_i - \langle  \oR_i \rangle)^2 -
\lam(I^\al_i - \langle  \oI_i\rangle )^2}~~,
                                          \label{10}
\eeq
where $D$ is a normalization factor defined such that $\langle 1\rangle =1$,
and $\lam$ is a measure of the inverse
width.
The quantities $\oR_i$ and $\oI_i$ are the mean values of $R_i$ and $I_i$
defined along a curve of length $N$ steps.\footnote{Note that this implies
that the mean value in a point is assumed to be equal to the mean value
computed along all curves of length $N$. Thus the mean values can
depend on $N$.}
Thus, eq. \rf{10} is relevant for a 3-dimensional world which is an
$N^3$ lattice.

\item[(ii)] I assume that the mean values are isotropic, i.e.
$\langle  \oR_1\rangle = \langle  \oR_2\rangle = \langle  \oR_3\rangle$ and
$\langle  \oI_1\rangle = \langle  \oI_2\rangle = \langle  \oI_3\rangle $.
\end{itemize}

Assumption (i) is certainly the simplest way of implementing lack of
correlation of the gradient vectors over distances compatible
with the horizon scale, whereas
assumption (ii) is natural as there is no reason to expect any
preferred direction.

It should be noted that the distribution \rf{10} factorizes into an
$I$-part and an $R$-part.
Thus, for any expectation value consisting of $I$'s and $R$'s one has
factorization,
\beq
\langle \oR^{\al_{1}}_{i_{1}} \ldots \oR^{\al_{n}}_{i_{n}}
{}~~\oI^{\beta_{1}}_{j_{1}} \ldots  \oI^{\beta_{m}}_{j_{m}} \rangle
= \langle \oR^{\al_{1}}_{i_{1}} \ldots
\oR^{\al_{n}}_{i_{n}}\rangle \langle \oI^{\beta_{1}}_{j_{1}}
\ldots \oI^{\beta_{m}}_{j_{m}}   \rangle~~.                     \label{11}
\eeq
This property turns out to be very useful in computing the higher moments.

The expectation value of a component $B_i$ of the
magnetic field can now easily be found.
{}From the expression \rf{2} one finds that
\beq
B_i  =  \frac{1}{2} \ve_{ijk} F_{jk} = - i~\ve_{ijk} V^\dagger_j V_k
 =  2 \ve_{ijk} R_j I_k~~.                    \label{12}
\eeq
Thus
\beq
\oB_j  =  \frac{1}{N} \sum^N_{i=1} B^{\al_{i}}_j
 =  2 \ve_{jlk} \frac{1}{N} \sum^N_{i=1} R^{\al_{i}}_l I^{\al_{i}}_k~~.
                                                \label{13}
\eeq
Hence
\bea
\langle \oB_j \rangle & = & \frac{2}{N} \ve_{jlk} \langle
\sum^N_{i=1}~R^{\al_{i}}_l
                I^{\al_{k}}_k \rangle \nn
         & = &  \frac{2}{N} \ve_{jlk} \langle \sum^N_{i=1}~
              (R^{\al_{i}}_l - \langle \oR_l \rangle )
                (I^{\al_{i}}_k - \langle \oI_k \rangle)
 + N \langle \oR_l \rangle \langle \oI_k \rangle \rangle~~.
                                    \label{14}
\eea
Now, due to the factorization \rf{11}, the first term on the right-hand
side of the last Eq. \rf{14} vanishes\footnote{Because
$\langle R^{\al_{i}}_l - \langle  \oR_l \rangle\rangle = \langle I^{\al_{i}}_k
- \langle \oI_k \rangle\rangle = 0$
for symmetry reasons.},
and hence
\beq
\langle \oB_j \rangle = 2 \ve_{jlk} \langle \oR_l\rangle\langle \oI_k\rangle =
\ve_{jlk} \left(
\langle  \oR_l\rangle\langle \oI_k\rangle - \langle \oR_k\rangle\langle
\oI_l\rangle \right) = 0       \label{15}
\eeq
because of the isotropy assumption (ii).
Consequently {\em the mean value of the magnetic field vanishes}.

The second order moment is given by
\bea
\langle \oB^2_i \rangle & = & \frac{4}{N^2} \sum_{\al \beta} \langle R^\al
R^\beta \cdot
       I^\al I^\beta - R^\al I^\beta \cdot I^\al R^\beta \rangle \nn
& = &   \frac{4}{N^2}  \sum_{\al \beta} \left\{ \langle R^\al R^\beta \rangle
       \langle I^\al I^\beta\rangle - \langle R^\al_i R^\beta_j\rangle
\langle  I^\beta_i  I^\al_j \rangle \right\}~~,
                                          \label{16}
\eea
using the factorization \rf{11}.
Now
\bea
\langle R^\al_i R^\beta_j \rangle& =& \prod_\g \int~ \frac{d^3R^\g}{D}~R^\al_i
R^\beta_j
e^{- \lam (R^\g - \langle \oR\rangle)^2} \nn
&=& \frac{1}{2 \lam} \del_{ij} \del^{\al \beta} + \prod_\g \int
\frac{d^3R^\g}{D} \left[ \langle \oR_i\rangle R^\beta_j +
\langle \oR_j\rangle R^\al_i
 - \langle \oR_i\rangle \langle \oR_j\rangle \right]
e^{- \lam (R^\g - \langle \oR \rangle)^2}~,                \label{17}
\eea
and similarly for $\langle I^\al_i I^\beta_j\rangle$.
Further we have e.g.
\bea
\prod_\g ~\int~ \frac{d^3 R^\g}{D}~ R^\beta_j e^{- \lam (R^\g - \langle
\oR\rangle)^2}
&=&  \prod_\g ~\int~ \frac{d^3 R^\g}{D}~ (R^\beta_j - \langle \oR_j\rangle)
e^{- \lam (R^\g - \langle \oR\rangle)^2} + \langle \oR_j\rangle \nn
&=& \langle \oR_j\rangle~~,                                        \label{18}
\eea
i.e., the mean value in a given arbitrary point $\beta$ on the lattice is
equal to the mean value computed over all curves.
Using Eqs. \rf{17} and \rf{18} in Eq. \rf{16} we get
\beq
\langle \oB^2_i\rangle = \frac{4}{N^2} \sum_\al \left( \frac{3}{2 \lam^2} +
\frac{1}{\lam}
( \langle \oI\rangle^2 + \langle \oR\rangle^2) \right)
 + \frac{4}{N^2} \sum_{\al \beta}
\left( \langle \oR\rangle^2  \langle \oI\rangle^2 -
(\langle  \oR\rangle \langle \oI\rangle)^2 \right)~~.         \label{19}
\eeq
The first term is $\cO ({N}/{N^2}) = \cO ({1}/{N})$.
The last term, being the square of the mean value, actually vanishes
because of isotropy. Thus we conclude that the {\em rms} value of the magnetic
field scales like
\beq
\sqrt{\langle \oB^2_i\rangle} = \frac{2}{N} \sqrt{\sum_\al \left( \frac{3}{2
\lam^2}
+ \frac{1}{\lam} ( \langle \oI\rangle^2 + \langle
\oR\rangle^2 ) \right)} \sim
\cO (\frac{1}{\sqrt{N}}) ~~.                           \label{22}
\eeq

The reason for this slow decrease is the fact that isotropy prevents the
mean value from entering in $\langle \oB_i\rangle$ and $\langle \oB^2\rangle$,
and that the correlations
of the gradient vectors are of short range.

\section{Consequences of the electroweak magnetic field}
\def\brms{B_{\rm rms}}
\def\rhob{\rho_B}
\def\rhog{\rho_\gamma}
Let us now assume that at the time of the electroweak phase
transition, a magnetic field with a coherence length $\xi_0$ is generated,
with a scaling as given by Eq. \rf{22}.
Such a field  evolves according to usual magnetohydrodynamics
\beq
{\pa {\bf B}\over \pa t}=\nabla \times ({\bf v}\times {\bf B})
-\sigma^{-1}\nabla\times\nabla\times {\bf B},
\eeq
where the conductivity $\sigma\sim\infty$ in the early universe.
Accordingly, the field is then imprinted on the charged
 plasma\footnote{There is a possible caveat here: in very large magnetic fields
such as considered
here the velocity of the plasma might depend on the background magnetic field.
In the following I will neglect this.}.

 At later times the original coherence
length is redshifted by the expansion according to
\begin{equation}
\xi(t)={a(t)\over a_0}\xi_0~~~~. \label{31}
\end{equation}
The frozen--in magnetic field is also redshifted by the expansion of the
universe. Thus
at later times at the physical distance scale $L=N\xi$ one finds,
\begin{equation}
\brms (t,L) =B_0\left( {a_0\over a(t)}\right)^2{1\over\sqrt{N}}=B_0
\left({t_0\over t_*}\right)^{3\over 4}\left({t_*\over
t}\right)\left({\xi_0\over L}\right)^{1\over 2},            \label{32}
\end{equation}
where $T_0^2t_0=0.301\; M_P/\sqrt{g_*(T_0)}$ with $g_*$ the effective number
of degrees of freedom, and $t_*\simeq 1.4\times 10^3 (\Omega_0h^2)^{-2}$ yrs is
the time when the universe becomes matter dominated; for definiteness, we
shall adopt the the value $\Omega_0h^2=0.4$, which is the upper limit
allowed by the age of the universe.

It is not obvious what the coherence length $\xi$ actually is. It is likely
that it is macroscopic and much larger than the interparticle separation.
Let me however assume for brevity that $B_0\simeq
10^{24}$ G and $\xi\simeq 1/T$.
It is then easy to find from Eq. \rf{32} the size
of the cosmological field today.
Taking $t=1.5\times 10^{10}$ yrs and $L=100$ kpc (corresponding to $N=1.0\times
10^{24}$), we find that today the
electroweak magnetic field at the scale of intergalactic distances is
\begin{equation}
\brms =4\times 10^{-19}\ {\rm G}~~.       \label{33}
\end{equation}
This seems to be exactly what is required for the numerical dynamo simulations
to produce the observed galactic magnetic fields of the order $10^{-6}$ G. The
inherent uncertainties in the estimate \rf{33} are: the value of $\Omega_0h^2$
used for computing $t_*$; the time at which the magnetic field froze, or $T_0$;
the actual value of the field $B_0$. Therefore one should view \rf{33} as an
order--of--magnitude estimate only.

We should also check what other possible cosmological consequences the
existence of the random magnetic field, Eq. \rf{32}, may have. Let us first
note that the energy density $\rhob$ in the {\em rms} field is very small.
In the
radiation dominated era we find that the energy density within a horizon volume
$V$ is
\begin{equation}
\rhob ={1\over 2V}\int_0^{r_H} d^3{\bf r}\brms^2={3\over 4}
B_0^2\left({T\over T_0}\right )^4{1\over r_HT}~~.     \label{34}
\end{equation}
The
horizon distance is $r_H=2t$ so that $\rhob \sim T^5/M_P\ll \rhog$, and
the magnetic field contribution to the total energy density is negligible.

In principle, magnetic fields could modify primordial nucleosynthesis, as
discussed in \cite{cheng}. However, the electroweak magnetic field at
the time of nucleosynthesis and at the horizon scale
is only
$\brms\simeq 1500$ G which is far too small to give rise to any modifications
of the Big Bang nucleosynthesis.

\section{Ferromagnetic universe}

There is also another, more exotic possibility
for producing magnetic fields, which is based on
the observation that, due to quantum fluctuations,
the Yang--Mills vacuum is unstable in a large enough background magnetic
field \cite{savvidy}. There are indications
from lattice calculations that
this is a non-perturbative result \cite{tw}.
Such magnetic field fluctuations in
the early universe could be sufficient to trigger the phase transition to
a new, ferromagnet--like ground state with a magnetic field made
permanent by the  charged plasma. In this scenario the primordial
field is thus generated as a non-perturbative quantum effect.

The new vacuum  results provided
the $\beta$--function has a
Landau singularity:
\beq
\left\vert~\int^{\infty}_g{dx\over\beta (x)}~\right\vert <\infty.
\eeq
Then the effective Lagrangian has a minimum away from the perturbative
ground state ${\rm Tr}\; F^2=0$, given by
\beq
\frac 12g^2{\rm Tr}\; F^2_{\mu\nu}\vert_{\rm min}=\Lambda^4,
\eeq
where $\Lambda$ is the renormalization group invariant scale
\beq
\Lambda=\mu\exp\left(-\int_\infty^g{dx\over\beta (x)}\right),
\eeq
where $\mu$ is a subtraction point associated with the definition of $g$.

The condition for the minimum can be realized in many ways. One of them is
 a constant non--abelian magnetic field $B^a_i=\epsilon_{ijk}F^a_{jk}$
with a non--zero component only in one direction in the group space,
and with a length given by
\beq
g\sqrt{B^aB^a}=\Lambda^2.
\eeq

Consider now SU(N) at the one--loop level. We then have
the one--loop, zero temperature  effective energy for a constant
background non--abelian magnetic field which in pure SU(N) theory reads
\cite{savvidy}
\beq
V(B)=\frac 12 B^2+ \frac{11N}{96\pi^2}g^2B^2\left(\ln {gB\over\mu^2}-
\frac 12\right)              \label{101}
\eeq
with a minimum at
\beq
gB_{\rm min}=\mu^2\exp \left(-{48\pi^2\over 11Ng^2}\right)  \label{102}
\eeq
and $V_{\rm min}\equiv V(B_{\rm min})=-0.029(gB_{\rm min})^2$.
Thus the ground state (the Savvidy vacuum)
has a non--zero  non--abelian magnetic field, the magnitude
of which is exponentially suppressed relative to the renormalization
scale, or the typical momentum scale of the system. Thus, for example,
for SU(2)$_{\rm L}$ at the electroweak scale the vacuum magnetic field
would be very small.
In the early
universe, however, where possibly a grand unified symmetry is
valid, the exponential suppression is less severe. It is also
attenuated by the running of the coupling constant. For a set of
representative
numbers, one might consider a (susy) SU(5) model with $\gut\simeq 1/25$
and $T_{\rm GUT}\simeq 10^{15}$ GeV, as in
the supersymmetric Standard Model. This yields $B\simeq 5\times 10^{-8}\mu^2$,
which turns out to be a magnitude which is relevant for the dynamo mechanism.

In the early universe the effective energy picks up thermal corrections
from fermion\-ic, gauge boson, and Higgs boson loops. In SU(2) these are
obtained by summing the Boltzmann factors $\exp(-\beta E_n)$ for the
oscillator modes
\beq
E_n^2=p^2+2gB(n+\frac 12)+2gBS_3+m^2(T), \label{1021}
\eeq
where $S_3=\pm 1/2~(\pm 1)$ for fermions (vectors bosons).
In Eq. \rf{1021} I have included the thermally induced mass $m(T)\sim gT$,
 corresponding to
a ring summation of the relevant diagrams. Numerically, the effect of
the thermal mass turns out to be very important.

The detailed form of
the thermal correction depends on the actual model, but we may take our
cue from the SU(2) one--loop calculation, which for the
fermionic and scalar cases can be extracted from
the real--time QED calculation in \cite{thermal}. The result is
\bea
\delta V_T^f&=&\frac
{(gB)^2}{4\pi^2}\sum_{l=1}^{\infty}(-1)^{l+1}\int_0^{\infty}
\frac{dx}{x^3}e^{-K_l^a(x)}\left[\; x{\rm coth}(x)
-1\right], \nn
\delta V_T^s&=&\frac {(gB)^2}{8\pi^2}\sum_{l=1}^{\infty}\int_0^{\infty}
\frac{dx}{x^3}e^{-K_l^a(x)}\left[\; x{1\over
{\rm sinh}(x)}-1\right],
\label{3}
\eea
where the normalization is such that
the correction vanishes for zero field, and
\beq
K_l^a(x)={gBl^2\over 4xT^2}+{m_a^2x\over gB}
\eeq
where $a=f,~b$ stands for fermions or bosons.

For vector bosons there is the added complication that there exists a
negative, unstable mode, which gives rise to an imaginary part\footnote{
Physically the  imaginary part is an indicator that the
vacuum also contains vector particles \cite{no}.}.
At high temperatures the instability is absent for fields such that
$gB<m^2(T)$, which is the case we are interested in here, so that no
regulation of the unstable $n=0,~S_3=-1$ mode is needed.
Thus  we find \cite{eo3}
\beq
\delta V_T^v=\frac {(gB)^2}{8\pi^2}\sum_{l=1}^{\infty}\int_0^{\infty}
\frac{dx}{x^3} e^{-K_l^b(x)}\left[\;
x{{\rm cosh}(2x)\over {\rm sinh}(x)}-1\right].  \label{104}
\eeq


At high temperature, the bosonic contributions are more important than
the fermionic ones. When $B\ll T^2\simeq m^2(T)$, we find numerically that
 $\delta V_T^v\simeq 0.016\times (gB)^2$.
This gives rise to a small correction to the magnitude of the field at
the minimum as obtained from Eq. \rf{102}. We may thus conclude that the
Savvidy vacuum  exists for all $T$.

The transition
to this new ferromagnet-like vacuum is triggered by local fluctuations.
Charged particles in
the primeval plasma  generate  current ${\bf j}=\nabla \times {\bf B}$.
The typical interparticle distance is $L\sim 1/T$ and a typical curl
goes like $1/L$
 so that $B\sim jL$ where
$j$ is like charge density with one charge in the  volume
$L^3$. Thus the Maxwell equations imply that $B\sim 1/L^2=T^2$,
indicating that the creation of the
Savvidy vacuum can take place locally.
A constant  non--abelian magnetic field, given by Eq. \rf{102},
is then imprinted on the
plasma of particles carrying the relevant charges.
The  Maxwell
magnetic field $B_{em}$ is a projection in the space of non--abelian
magnetic
fields, and we take it to be of the  size comparable to $B$ in Eq. \rf{105}.

The magnetic flux remains
conserved (recall that
the primodial plasma is an extremely good conductor), and we may
write
\beq
B(T)=g_{\rm GUT}^{-1}\mu^2\exp \left(-{48\pi^2\over 11Ng^2}\right)
\left({T^2\over \mu^2}\right)\simeq
3\times 10^{42}\G\left({a(t_{\rm GUT})\over a(t)}\right)^2, \label{105}
\eeq
where $\mu\simeq T$ and
$a(t)$ is the scale factor of the universe, and the last figure is
for susy SU(5). This expression is valid because
 the energy $E$ of the vacuum is redshifted
 by $1/a(t)$. Now in the minimum
 $E$ is proportional to $VB^2$, where $V$ is the volume.
Since $V$ is proportional to $a^3$, we
get $B\sim 1/a^2$.
Hence the magnetic energy per horizon is much less than the radiation
energy.

As time passes, the
universe undergoes a number of phase transitions. Each of these correspond
to new types of ferromagnetic vacua,
which in general have decreasing field strengths.
However, the original GUT vacuum has existed for a time that
is long enough for the plasma to interact with the vacuum field $B$
given by Eq. \rf{105}. This interaction does
not allow for the GUT flux to decrease
once it has been created since it has become a feature of the plasma, which
conserves the flux in the sense that the magnetic lines of force are
frozen into the fluid.   Even if
the original field is suddenly
removed by the creation of a new
vacuum, the magnetic field will
survive in a perfect conductor (see e.g. pp.
186-189 in ref. \cite{ll}).

{}From Eq. \rf{105} we find that the Maxwell magnetic field at
$t_{now}\simeq 10^{10}$ yr is given
by $B_{now}\simeq 3\times 10^{42}\G (t_{\rm GUT}/t_*)(t_*/t_{now})^{4/3}
\simeq 10^{-14}\G$. Such a magnetic field appears comparable
to what is needed for the seed field in galactic dynamo models.
Note also that at nucleosynthesis one obtains $B\sim 10^4$ G, which is well
below the nucleosynthesis bound on magnetic fields \cite{cheng}.

\section{Discussion}

The GUT causal domain $l_0$ has  today the size of only about
1 m.
Obviously, during the course of the evolution of the universe, domains
with magnetic fields pointing to different directions have come into
contact with each other. One might think that
this results in domain walls. However, here it is important that the
magnetic flux lines follow the plasma particles and cease
to be homogenous. If inside each GUT
horizon  a magnetic field line at a certain time passes through
two plasma particles, then this is true at any later time. The
magnetic field thus "aligns" with the plasma. When
two horizon  bubbles collide, the two plasmas rearrange
and become one plasma, and the same is true for the magnetic field
lines, which are part of the new plasma \cite{eo4}. The  field
"realigns" with the "new" plasma and the root mean square of
the magnetic field  remains of the same order as before.
Because there are no domains, no random walk factor appears at large
distances.

 In this argument it is important that
$\brms$ is much smaller than the square of the rms momentum, since
otherwise the electrical conductivity would depend on  $B$. This condition
implies that the radius of curvature
of a typical  plasma particle is very large compared to the mean free path,
or that the magnetic energy is much less than the kinetic energy of the
plasma. This certainly is the case in the ferromagnetic universe model.

The size of the field is determined by the scale at which the
ferromagnet vacuum is created, and the earlier this happens, the bigger
the field. If there is a period of cosmic inflation, then the
relevant field would be created after reheating. If the
reheating temperature is comparable to GUT scales, the
strength of the magnetic field would be given as in \rf{105}, with
interesting consequences for the formation of galactic magnetic fields.
In this scenario the primordial seed field is thus a relic from the GUT era.

If the origin of the magnetic field is the  electroweak phase transition,
the situation might be different because at  scales $1/T$ the magnetic energy
equals to the energy in radiation. In such fields the plasma might be
trapped by the field, rather than the field being imprinted on the moving
plasma. Very likely this would result in a domain structure, but in the
absence of any true dynamical calculation, the details must remain unclear.
It nevertheless seems  clear that  a primordial magnetic field
would have many intriguing consequences, some of which might actually
be observable.
It would of course be of great interest to detect this relic field
directly in the intergalactic space.

\vskip1truecm\noindent
{\Large\bf Acknowledgements}
\vskip0.5truecm
I wish to thank Poul Olesen for many enjoyable discussions on primordial
magnetic fields.

\end{document}